\documentclass[prb,twocolumn,amsmath,amssymb,showpacs]{revtex4}

\usepackage{graphicx}
\usepackage{dcolumn}
\usepackage{bm}
\usepackage{mathrsfs}
\usepackage{appendix}
\usepackage{bbm}
\usepackage{color}
\def \Ups {\Omega }
\def \vk {{\bf k}}

\def \ni {{n_i}}

\def \ein {\epsilon^{\infty }}
\def \vH {{\bf H}}

\def \vS {{\bf S}}
\def \vR {{\bf R}}
\def \ve {{\bf e}}
\def \vQ {{\bf Q}}

\def \vU {{\bf T}}

\def \vh {{\bf h}}
\def \vv {{\bf v}}
\def \vm {{\bf m}}

\def \mb {\mu_{\rm B}}
\def \mal {{\bar \alpha}}
\def \vu  {{\bf u}}

\def \pin {{\bf P}^{\rm ind} }
\def \paf {{\bf P}^{\rm ind}_{\rm AF} }

\def \psc {{\bf P}^{\rm SC}}
\def \pms {{\bf P}^{\rm MS}}

\def \cV {{\cal V}}

\def \xp {{\bf x}^{\prime }}
\def \yp {{\bf y}^{\prime }}
\def \zp {{\bf z}^{\prime }}

\def \zz {{\bf z}}

\def \vM {{\bf M}}
\def \vP {{\bf P}}
\def \vW {{\bf W}}
\def \BF {{\rm BiFeO$_3$} }
\def \BP {{\rm BiFeO$_3$}}

\def \TN {T_{\rm N}}
\def \vE {{\bf E}}

\def \vk {{\bf k}}
\def \vv {{\bf v}}

\def \xq {x^{\prime}}
\def \yq {y^{\prime}}
\def \zq {z^{\prime}}
\def \ds {\displaystyle }

\def \cvMa { \langle n \vert \vM \vert 0 \rangle }

\def \cvPa {\langle n \vert \vP^{{\rm ind}} \vert 0 \rangle }

\def \vh {{\bf h}}
\def \vn {{\bf n}}
\def \vz {{\bf z}}
\def \vx {{\bf x}}
\def \vy {{\bf y}}

\def \mB {\mu_{\rm B}}

\begin{document}

\title{Spin-Induced Polarizations and  Non-Reciprocal Directional Dichroism of the Room-Temperature Multiferroic BiFeO$_3$\footnote{Copyright notice: This manuscript has been authored by UT-Battelle, LLC under Contract No. DE-AC05-00OR22725 with the U.S. Department of Energy. The United States Government retains and the publisher, by accepting the article for publication, acknowledges that the United States Government retains a non-exclusive, paid-up, irrevocable, world-wide license to publish or reproduce the published form of this manuscript, or allow others to do so, for United States Government purposes. The Department of Energy will provide public access to these results of federally sponsored research in accordance with the DOE Public Access Plan (http://energy.gov/downloads/doe-public-access-plan).}}

\author{Randy S. Fishman$^1$, Jun Hee Lee$^1$, S\'andor Bord\'acs$^2$, Istv\'an K\'ezsm\'arki$^2$, Urmas Nagel$^3$, and Toomas R\~o\~om$^3$}

\affiliation{$^1$Materials Science and Technology Division, Oak Ridge National Laboratory, Oak Ridge, Tennessee 37831, USA}
\affiliation{$^2$Department of Physics, Budapest University of Technology and Economics and MTA-BME Lend\"ulet Magneto-optical Spectroscopy
Research Group, 1111 Budapest, Hungary}
\affiliation{$^3$National Institute of Chemical Physics and Biophysics, Akademia tee 23, 12618 Tallinn, Estonia}

\date{\today}

\begin{abstract}

A microscopic model for the room-temperature multiferroic \BF that includes two Dzyaloshinskii-Moriya interactions and single-ion anisotropy along the ferroelectric polarization
predicts both the zero-field spectroscopic modes as well as their splitting and evolution in a magnetic field.  
Due to simultaneously broken time-reversal and spatial-inversion symmetries, the absorption of light changes as the 
magnetic field or the direction of light propagation is reversed.  We discuss three physical mechanisms that may
contribute to this absorption asymmetry known as non-reciprocal directional dichroism:  the spin current, magnetostriction, and single-ion anisotropy.   
We conclude that the non-reciprocal directional dichroism in \BF is dominated by the spin-current polarization and is insensitive to the magnetostriction and easy-axis anisotropy.
With three independent spin-current parameters, our model accurately describes the non-reciprocal directional dichroism observed for magnetic field along $[1,-1,0]$.
Since some modes are almost transparent to light traveling in one direction but opaque for light traveling in the opposite direction, 
\BF can be used as a room-temperature optical diode at certain frequencies in the GHz to THz range.  Our work demonstrates that
an analysis of the non-reciprocal directional dichroism spectra based on an effective spin model supplemented by first-principles calculations
can produce a quantitative microscopic theory of the magnetoelectric couplings in multiferroic materials.

\end{abstract}


\maketitle

\section{Introduction}

\BF is the only material known to exhibit multiferroic behavior at room temperature.  
Because its ferroelectric (FE) transition temperature \cite{teague70}
$T_c \approx 1100$ K is significantly higher than its N\'eel transition temperature \cite{sosnowska82} $\TN \approx 640$ K, \BF is a type I
multiferroic.  Although the non-magnetic FE polarization \cite{lebeugle07} $P^{\rm FE} \approx 90\, \mu$C/cm$^2$ is
much larger than the magnetic contribution \cite{kadomtseva04, tokunaga10, park11, lee13} induced by the distorted spin cycloid 
\cite{sosnowska82, lebeugle08, rama11a, herrero10, sosnowska11}, the magnetic domain 
distribution of \BF can be manipulated by an applied electric field \cite{lebeugle08, slee08}.  

A great deal has been learned about \BF since the first single crystals became available for inelastic neutron scattering \cite{jeong12, matsuda12, xu12}, 
Raman scattering \cite{cazayous08, rov10}, and THz spectroscopy \cite{talbayev11, nagel13} measurements.
It is now understood that two sets of interactions control the cycloid of \BP :  two Dzyaloshinskii-Moriya (DM) interactions produced by broken inversion symmetry
and a single-ion anisotropy \cite{sosnowska95} (ANI) $K$ along the direction of the FE polarization $\vP^{\rm FE}$.  
Whereas the DM interaction \cite{sosnowska82} $D_1$ perpendicular to $\vP^{\rm FE}$ is responsible for the long 62 nm cycloidal period, 
the DM interaction \cite{kadomtseva04, ed05, pyatakov09, ohoyama11} $D_2$ along $\vP^{\rm FE}$
is responsible for a small cycloidal tilt \cite{pyatakov09}.
Above the critical magnetic field $H_c$, the cycloidal tilt develops into the weak ferromagnetic (FM) moment \cite{tokunaga10, park11, rama11b}
of a G-type antiferromagnet (AF) that is isosymmetrically canted by an antiferrodistortive rotation 
(R$_4^+$[1,1,1]) of the $R3c$ structure \cite{ed05}.

Inelastic neutron scattering measurements \cite{jeong12, matsuda12, xu12} were used to extract the AF nearest- and 
next-nearest neighbor exchange interactions \cite{Hes} $J_1=-5.32$ meV and $J_2=-0.24$ meV between the $S=5/2$ Fe$^{3+}$ spins
on the pseudo-cubic unit cell sketched in Fig.1(a) with lattice constant $a = 3.96 \AA $.
However, those measurements lacked the sensitivity to resolve the ordering wavevectors on either side of 
the G-type AF wavevector $\vQ_0 = (2\pi /a)[0.5,0.5,0.5]$ at $(2\pi /a) [0.5 \pm \delta ,0.5, 0.5\mp \delta ]$, 
where $\delta \approx 0.0045$ is inversely proportional to the cycloidal period $a/(\sqrt{2} \delta )$.
Recent neutron scattering measurements \cite{jeong14} with higher precision were able to distinguish the two cycloidal ordering
wavevectors and found that \cite{Hes} $D_1 = 0.18$ meV and $K=0.0039$ meV.  But even those
measurements lacked the precision to obtain $D_2$, which was set to zero.

By contrast, the frequencies of the spin-wave (SW) modes at the ordering wavevector $\vQ $ can be precisely measured with Raman scattering \cite{cazayous08} and 
THz spectroscopy \cite{talbayev11, nagel13}.  The parameters $K$, $D_1$, and $D_2$ were
estimated by fitting the frequencies \cite{fishman13a} of the four observed zero-field THz modes.  With no remaining adjustable parameters,
that same model predicted \cite{fishman13b} the evolution and activation of the THz modes \cite{nagel13} in a magnetic field 
along $[0,0,1]$.

We now use this microscopic model to predict the asymmetry $\Delta \alpha (\omega )$ in the absorption 
$\alpha (\omega )$ of light when the direction of the magnetic field or, equivalently, the direction of light propagation is reversed.  
Called non-reciprocal directional dichroism (NDD), absorption asymmetry was first observed by Hopfield and Thomas \cite{hop60} over 50 years ago in CdS.
Much more recently, the precise symmetry requirements for NDD in magnetic materials were systematically investigated by Szaller {\em et al.} [\onlinecite{szaller13}].
Strong NDD is expected for the spin excitations of multiferroic materials when both time reversal and spatial
inversion symmetries are broken by the spin state.  Both the magnetic and electric components of THz radiation can excite SWs in multiferroic materials.  
The NDD exhibited by simultaneously electric- and magnetic-dipole active
excitations has been extensively studied in Ba$_2$CoGe$_2$O$_7$ [\onlinecite{kezsmarki11, bordacs12, miyahara11, kezsmarki14}], 
Sr$_2 $CoSi$_2$O$_7$ [\onlinecite{kezsmarki14}], Ca$_2$CoSi$_2$O$_7$ [\onlinecite{kezsmarki14}], Gd$_{0.5}$Tb$_{0.5}$MnO$_3$ [\onlinecite{takahashi13}], 
and Eu$_{0.55}$Y$_{0.45}$MnO$_3$ [\onlinecite{takahashi12}].

Because the cycloidal spin state is produced by
the competition between DM, exchange, and ANI interactions, three distinct physical mechanisms can produce NDD in \BP :
the spin current (SC) driven by the DM interactions, magnetostriction (MS) or the electric-field induced changes in the exchange interactions,
and the electric-field induced changes in the ANI.   
Remarkably, the dynamical magnetoelectric coupling governing the NDD in \BF is dominated by the two sets of SC
polarizations associated with $D_1$ and $D_2$.  Qualitatively, the SC dominates the magnetoelectric coupling in \BF because
spin fluctuations $\delta \vS_i$ 
are transverse to the almost collinear, cycloidal spin state $\langle \vS_i \rangle $.  Since $\delta \vS_i \times \langle \vS_j \rangle \ne 0$ but
$\delta \vS_i \cdot \langle \vS_j \rangle \approx 0$
(for nearby sites $i$ and $j$) and the ANI is extremely weak, spin fluctuations more strongly affect the SC-induced polarization than 
the MS- and ANI-induced polarizations.  

As a fraction of the total light absorption at a given wavelength, NDD is most pronounced for a mode with fluctuations out of the cycloidal plane at 15.5 cm$^{-1}$.
At this wavenumber, \BF is almost transparent for light traveling in one direction but opaque for light traveling in the
opposite direction.  Therefore, \BF can be used as an optical diode that operates up to room temperature.

\begin{figure}
\includegraphics[width=8.5cm]{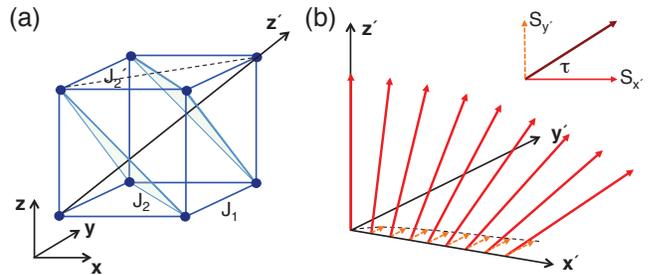}
\caption{(Color online) (a) The exchange interactions $J_1$ and $J_2$ on the pseudo-cubic lattice for \BF with Fe$^{3+}$ ions at the corners of the cube and
two hexagonal layers shown.  Due to the rhombohedral
distortion along $\zp $, $J_2^{\prime }$ and $J_2$ may be slightly different.  (b) The spin state in zero magnetic field, both with electric polarization along $\zp $.  
The canting of the spins out of the $\{ \xq ,\zq \}$ plane is indicated by the angle $\tau $ in the inset.  The variation of the canted component along $\xq $
is also shown by the dashed line in (b).}
\end{figure}

Despite the success of our model describing the NDD for magnetic field along $[1,-1,0]$, several questions remain open.
Although our model predicts NDD to be absent for light propagating along $\vk = [0,0,1]$, a static magnetic field along 
$[\eta ,\eta ,\kappa ]$, and THz electric-field orientation $\ve =[1,1,0]$ or $[1,-1,0]$, weak NDD has been observed for a magnetic field along $[1,1,0]$ under those conditions.  
An optical misalignment of the THz electric- and magnetic-field vectors $\ve $ and $\vh =\vk \times \ve $ may be responsible for this effect.
In addition, the mean absorption $\mal (\omega)$ (the absorption $\alpha (\omega )$ averaged over 
positive and negative magnetic fields) is not as accurately predicted by our model as the NDD.

This paper complements a recent work \cite{istun} that presents detailed experimental results for 
both the individual absorptions and the NDD.  We have divided this paper into six sections.
Section II presents our microscopic model and Section III presents the predicted mode frequencies.
Section IV describes the three polarization mechanisms and presents results for the magnetization and polarization matrix elements,
with symmetry relations provided by Local Spin-Density Approximation (LSDA)+$U$ calculations.
Results for the NDD are presented in Section V.  Section VI contains a discussion and conclusion.
While Appendix A summarizes the experimental details,
Appendices B, C, and D treat the SC-, MS-, and ANI-induced polarizations, respectively.
For convenience, the unit vectors used in this paper are given in Table I.

\section{Microscopic model}

In a magnetic field $\vH =H\vm $, the spin state and SW excitations of \BF are evaluated from the microscopic Hamiltonian
\begin{eqnarray}
\label{Ham}
&&{\cal H} = -J_1\sum_{\langle i,j\rangle }\vS_i\cdot \vS_j -J_2\sum_{\langle i,j \rangle'} \vS_i\cdot \vS_j
\nonumber \\
&&+D_1\, \sum_{\langle i,j\rangle } (\zp \times {\bf e}_{i,j}/a) \cdot (\vS_i\times\vS_j) \nonumber \\
&& + D_2\, \sum_{\langle i,j\rangle } \, (-1)^\ni \,\zp \cdot  (\vS_i\times\vS_j)\nonumber \\
&& -K\sum_i (\zp \cdot \vS_i )^2
- 2\mb H \sum_i \vm \cdot \vS_i ,
\end{eqnarray}
where $\ve_{i,j } = a{\bf x}$, $a{\bf y}$, or $a{\bf z}$ connects $\vR_i$ with its nearest neighbor $\vR_j=\vR_i + \ve_{i,j}$.
Since the unit vector $\zp $ points along a cubic diagonal parallel to the FE polarization $\vP^{\rm FE}$,
the $D_1$ sum has the form proposed by Katsura {\em et al.} \cite{katsura05}.  The hexagonal 
layers normal to $\zp$ are separated by $c=a/\sqrt{3}$ and are labeled by the integer $\ni =\vR_i \cdot \zp /c$.  
Consequently, the $D_2$ sum alternates sign from one hexagonal layer to the next.
Notice that the local DM interactions $D_1\,(\zp \times \ve_{i,j}/a)$ and $D_2\,\zp $ are, respectively, perpendicular and parallel to $\zp $.

\begin{table}
\caption{\textbf{Unit vectors}}
\begin{ruledtabular}
\begin{tabular}{llc}
 $\vx $, $\vy $, $\vz $ & Pseudo-cubic laboratory reference frame \\
 $\zp =\zp_m$ & Orientation of the electric polarization $\vP^{{\rm FE}}$  \\
$$ & along one of the cubic diagonals \\
 $\xp $, $\yp $, $\zp $ & Cycloidal reference frame \\
 $\xp_m $, $\yp_m $, $\zp_m$ & Cycloidal reference frame for domain $m$\\
 $\vu $ & $\vx $, $\vy $, or $\vz $\\
 $\vm $ & Orientation of the static magnetic field \\
 $\vn_i $ & Local single-ion ANI axis \\
 $\vk $ & Direction of light propagation \\
 $\ve $ & Orientation of the THz electric field \\
 $\vh $ & Orientation of the THz magnetic field \\
\end{tabular}
\end{ruledtabular}
\end{table}

There are eight possible orientations for $\vP^{\rm FE} \parallel \zp $ along the four cubic diagonals.   
For every possible $\zp $, the three magnetic domains have different $\xp $ and $\yp $.  When $\zp =  [1,1,1]$
(all unit vectors in Table I are assumed normalized to 1), the possible 
orientations for the $\xq $ axis are $\xp_1 = [1,-1,0]$, $\xp_2 = [1,0,-1]$, and $\xp_3 = [0,1,-1]$ with corresponding
$\yp_m = \zp \times \xp_m $.  These three magnetic domains have cycloidal ordering wavevectors 
\begin{equation}
\vQ_m = \vQ_0 + \frac{2\sqrt{2} \pi \delta }{a} \xp_m .
\end{equation}
Hence, the ordering wavevectors for different domains are $\vQ_1 = (2\pi /a)[0.5+\delta ,0.5-\delta, 0.5]$,
$\vQ_2 = (2\pi /a)[0.5+\delta ,0.5, 0.5-\delta ]$, and $\vQ_3 = (2\pi /a)[0.5,0.5+\delta ,0.5-\delta ]$.
In terms of $\delta \ll 1$, the period of the cycloid in zero field is $a/(\sqrt{2} \delta )\approx 62$ nm. 

As mentioned above, the DM interactions $D_1$ and $D_2$ only couple nearest-neighbor sites.  In 
a previous formulation \cite{fishman13a, fishman13b} of this microscopic model, $D_1$ coupled next-neighbor sites
within the same hexagonal layer.   
Due to the very long cycloidal period $p \gg a$ of \BP, the equilibrium and dynamical properties of these two Hamiltonians are the same
up to errors of order $\delta^2 \approx 2\times 10^{-5}$.  
Specifically, earlier predictions for the SW mode frequencies \cite{fishman13a, fishman13b} 
and critical magnetic field \cite{fishman13c} are unchanged.  However, the earlier DM interaction $D_1$ is now multiplied by $\sqrt{2}$. 
Because the nearest-neighbor DM interactions are much larger than those between next-neighbor spins, 
the Hamiltonian above provides a close connection with recent first-principles calculations \cite{ed05, junun}.

Since the $D_1$ and $D_2$ terms in ${\cal H}$ depend only on $\zp $, ${\cal H}$ is independent of the magnetic domain.
For a specific domain $m$, the first SC term can be written $V^{{\rm SC}}_1 = \sqrt{2} D_1 N\, \yp \cdot \vU_1 $,
where
\begin{equation}
\label{NDDT1}
\vU_1 = \frac{1}{N} \sum_{\langle i,j \rangle^{\bf x}} \Bigl\{ \vS_i \times \vS_j \Bigr\}, \,\,\, m=1,2,
\end{equation} 
\begin{equation}
\label{NDDT3}
\vU_1 = \frac{1}{N} \sum_{\langle i,j \rangle^{\bf y}} \Bigl\{ \vS_i \times \vS_j \Bigr\}, \,\,\, m=3,
\end{equation} 
where $\langle i,j\rangle^{\bf u}$ is a sum over nearest neighbors with $\vR_j - \vR_i = a\vu $.
These relations assume that the spins on each hexagonal layer depend only on the integer $r = \sqrt{2}\xp \cdot \vR_i /a$.
So for domain 2, $\vS (\vR_i + a\vx ) =\vS (\vR_i -a\vz )$.
The cross products in Eqs.(\ref{NDDT1}) and (\ref{NDDT3}) couple spins with indices $r$ and $r+1$ on neighboring layers.

The second SC term $V_2^{{\rm SC}}$ proportional to $D_2$ can be written
$V_2^{{\rm SC}}= D_2 N \, \zp \cdot \vU_2 $, where
\begin{equation}
\label{NDDT4}
\vU_2 = \frac{1}{N} \ds\sum_{\langle i,j\rangle } (-1)^\ni \Bigl\{ \vS_i \times \vS_j \Bigr\}.
\end{equation}
Like $V_1^{{\rm SC}}$, $V_2^{{\rm SC}}$ also couples neighboring spins on neighboring layers. 

The nearest- and next-nearest neighbor exchange interactions \cite{Hes} $J_1=-5.32$ meV and $J_2=-0.24$ meV were obtained from recent 
inelastic neutron scattering measurements \cite{jeong12, matsuda12, xu12} between 5.5 meV and 72 meV.  
On the other hand, the small interactions $D_1$, $D_2$, and $K$
that control the cycloid can be obtained from THz spectroscopy
measurements \cite{talbayev11, fishman13a} below 5.5 meV (44.3 cm$^{-1}$) in zero magnetic field. 

We have neglected the broken spatial symmetry between the exchange interactions due to the rhombohedral distortion.  
While all $J_1$ interactions must remain the same due to the rotational $C_3$ symmetry about $\zp $, 
$J_2$ may reflect the rhombohedral distortion.  For example, next-nearest neighbors 
separated by $\vR_1 = a({\bf x}+{\bf y})$ and $\vR_2 = a({\bf x}-{\bf y})$ may experience slightly different exchange interactions,
denoted by $J_2^{\prime }$ and $J_2$ in Fig.1(a), because
$\vR_1\cdot \zp =2a/\sqrt{3}$ while $\vR_2 \cdot \zp =0$.  
However, based on the excellent agreement between theory and experiment for the 
mode frequencies reported in Section III and because $J_2$ is already so small, we expect this exchange anisotropy to have a very minor effect on the NDD.
 
For a given set of interaction parameters, the spin state of \BF is obtained by minimizing the energy 
$E=\langle {\cal H}\rangle $ over a set of variational parameters \cite{fishman13b}.
Fixing $\delta  = 1/q $, where $q\gg 1 $ is an integer, the energy $E$ is minimized over the variational parameters
on a unit cell with $q$ sites along $\xp $ and two hexagonal layers.  The spin state on layer $n$ is assumed to be identical to the spin state 
on layer $n+2$.  The wavevector parameter $\delta $ is determined 
as a function of field by an additional minimization loop over $q$.  In zero field, $\delta \approx 0.0045$ and $q=222$.
We verify that the corresponding spin state provides at least a metastable minimum of the energy $E$
by checking that the classical forces on each spin vanish.

Ignoring the cycloidal harmonics $C_{l > 1}$ produced by $D_2$ and $K$ but including the 
tilt  \cite{pyatakov09} $\tau $ produced by $D_2$, the spin state in zero field
can be approximated by 
\begin{eqnarray}
\label{syc1}
S_{\xq }(\vR )&=& S (-1)^{n+1} \cos \tau \sin (2\pi \delta r ), \\
\label{syc2}
S_{\yq }(\vR )&=& S \sin \tau \sin (2\pi \delta r ), \\
\label{syc3}
S_{\zq }(\vR )&=&S (-1)^{n+1} \cos (2\pi \delta r ).
\end{eqnarray}
This tilted cycloid is plotted in Fig.1(b).
Averages over this state are readily performed 
using $\langle {S_{i \xq}}^2 \rangle =(S^2 /2)\cos^2 \tau $, $\langle {S_{i \yq}}^2 \rangle =(S^2 /2) \sin^2 \tau $, 
and $\langle {S_{i\zq}}^2 \rangle = S^2/2$.   In zero field, averages over the tilted cycloid are fairly accurate
because \cite{fishman13a} even harmonics like $C_2$ vanish and $C_3\approx 5\times 10^{-3}$.
Corrections to the averages are then of order ${C_3}^2 \approx 2.5 \times 10^{-5}$.

For comparison, the spin state of the canted AF at zero field can be simply written in terms of the canting angle $\tau $
within the $\{\xq ,\yq ,\zq \}$ coordinate system as 
\begin{equation}
\label{CAF}
\vS_n = S [(-1)^{n+1}\cos \tau , \sin\tau ,0 ]
\end{equation}
on hexagonal layer $n$.  
Recall that \cite{fishman13a} $\sin \tau = S_0/S$ where $2\mB S_0 $ is the weak FM moment of the AF phase 
along $\yp $ above $H_c$.  Whereas susceptibility measurements \cite{tokunaga10} indicate that $S_0=0.015$,
a recent neutron-scattering study \cite{rama11b} suggests that $S_0 \sim 0.05$ equivalent to $\tau \sim 1^\circ $.
By contrast, LSDA+$U$ ($U=5$ eV) \cite{junun} gives $S_0=0.014$, in agreement with the former experimental result.
Note that $D_2=-2J_1S_0/S = -2J_1 \sin \tau $ is a linear function of $S_0$ and of $\sin \tau \approx \tau$.
 
 \begin{figure}
\includegraphics[width=8.3cm]{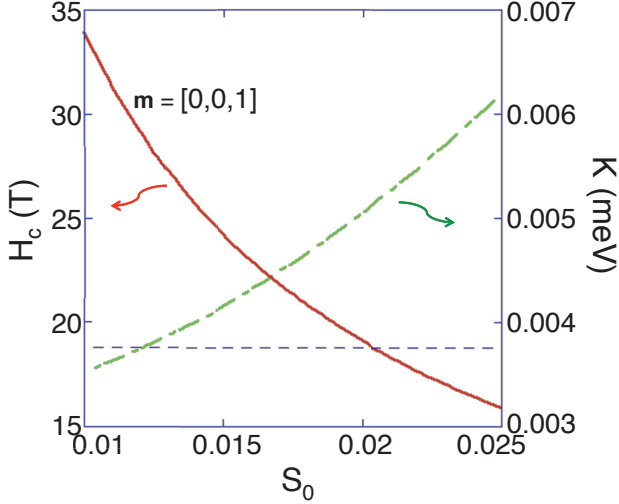}
\caption{(Color online) The predicted critical field (solid) versus $S_0$ 
for field orientation $\vm =[0,0,1]$.  The horizontal line is the experimental value \cite{tokunaga10} for $H_c$.
Also plotted is the single-ion ANI $K$ (dash-dot) versus $S_0$.
}
\end{figure}

We now adopt a different approach to estimate $D_2$.
The three parameters $D_1$, $D_2$, and $K$ are fixed by two conditions:
the period of the cycloid must match the measured period and and the frequencies of the
four predicted SW modes in zero field must match the measured frequencies \cite{fishman13a}.  
A third condition is provided by the dependence of the predicted critical field $H_c$ on $S_0$.
As shown in Fig.2, the measured critical field of 18.8 T for $\vm =[0,0,1]$ 
requires that \cite{Hes} $S_0=0.02$, corresponding to $\tau = 0.008$ or 0.45$^\circ $.
While $D_1 \approx 0.180$ meV is virtually independent of $S_0 $, $D_2$ linearly increases with
$S_0 $.  Figure 2 indicates that $K$ increases almost quadratically with $S_0$ from a value of $K=0.0031$ meV
when $S_0=0$.  Corresponding to $S_0=0.02$, we obtain $D_2=0.085$ meV and $K=0.0051$ meV.
A somewhat smaller value $K=0.0039$ meV was given in Ref.[\onlinecite{jeong14}], which took $D_2=0$.

With other parameters fixed and $\vm = [1,-1,0]$, a value of $D_2$ smaller than about 0.079 meV would stabilize
a different canted AF phase above $H_c$ with spins tilted above and below the $\{\xq ,\yq \}$ plane due to the dominant single-ion ANI.
Hence, the coplanar AF phase of Eq.(\ref{CAF}) is barely stabilized by the second DM interaction.

\section{Spectroscopic mode frequencies}

\begin{figure}
\includegraphics[width=8cm]{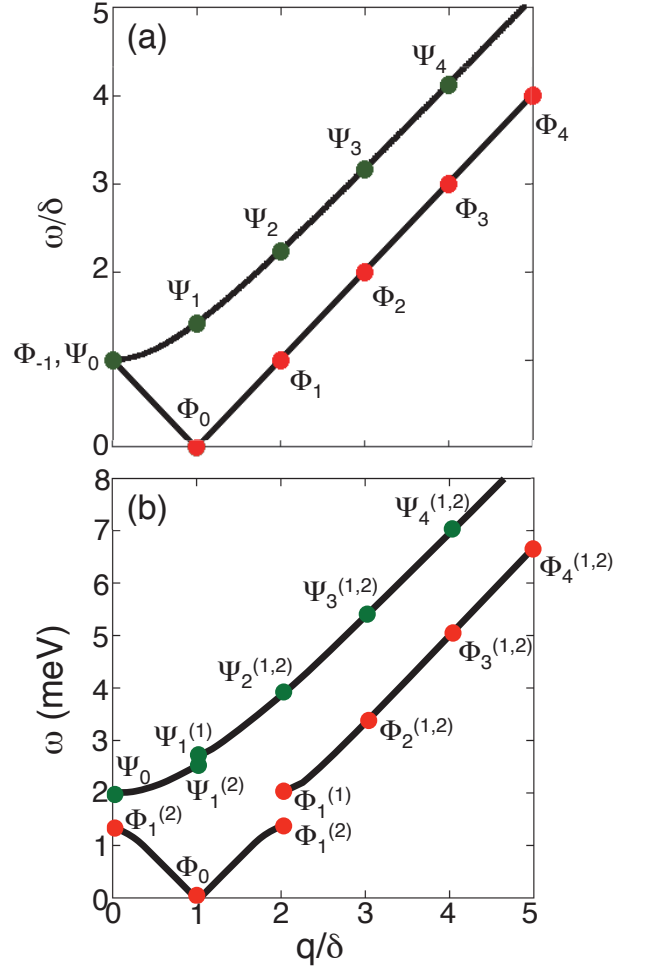}
\caption{(Color online) The mode spectra at multiples of the ordering wavevector in an extended zone scheme (a) without 
higher harmonics of the spin and (b) with higher harmonics for the predicted parameters of \BP .
}
\end{figure}

Using the parameters given above, $\cal{H}$ predicts the evolution of the modes with magnetic field \cite{nagel13, fishman13b, istun} 
for all orientations $\vm $.  SW modes at the ordering wavevector $\vQ $ can be labeled \cite{deSousa08} as in-cycloidal-plane 
$\Phi_m$ modes and out-of-cycloidal-plane $\Psi_m$ modes.  In an extended zone scheme, those mode frequencies are plotted
versus $q/\delta $ for wavevector $(2\pi /a) [0.5 + q, 0.5, 0.5 -q]$ in Fig.3(a).  For simplicity, $\Phi_m $ and $\Psi_m$ 
denote both the modes and their frequencies.  Neglecting higher spin harmonics, $\Phi_m = \vert m \vert \Psi_0$ and
$\Psi_m = \Psi_0 \sqrt{1+ m^2}$.  It follows that $\Phi_1 = \Psi_0$.

Higher harmonics generated by the tilt and ANI split 
each mode with $m \ge 1$ into two labeled $\Phi_m^{(1,2)}$ or $\Psi_m^{(1,2)}$.   For the predicted parameters of \BP , those 
modes are plotted versus wavevector in Fig.3(b).  While the $m=1$ modes are strongly affected by the spin harmonics,
the former mode scheme remains fairly accurate for $m >1$.   Because the splitting of the low-frequency modes was not
considered, recent Raman studies \cite{cazayous08, rov10} misidentified the observed modes with some out-of-plane modes mistaken for 
in-plane modes and vice versa.

Despite the substantial splitting of $\Phi_1^{(1)}$ and $\Phi_1^{(2)}$, $\Phi_1^{(1)}$ is only slightly larger than $\Psi_0$.  
The nearly degenerate $\Phi_1^{(1)}$ and $\Psi_0$ modes cannot be separated by THz measurements \cite{talbayev11, nagel13} in zero field.

In Fig.4, the predicted and measured \cite{missing} mode frequencies are plotted versus field for orientations $\vm = [0,0,1]$, $[1,1,0]$, and $[1,-1,0]$.
Experimental data was not available for the THz modes above $H_c$ for the last two field orientations.  The experimental results
for $\vm =[1,1,0]$ and $[1,-1,0]$ are presented here for the first time with experimental details summarized in Appendix A.  Because its frequency was too low,
$\Phi^{(2)}_1$ was not detected when $\vm = [1,1,0]$ and $[1,-1,0]$.
The predicted mode frequencies of the stable domain(s) are presented in the solid curves:  domain 1 for $\vm = [0,0,1]$ and $[1,1,0]$
and domains 2 and 3 for $\vm = [1,-1,0]$.  For $\vm = [0,0,1]$, the mode that dips below $\Phi_1^{(2)}$ arises from metastable domains 
2 and 3, as seen by the agreement with the dashed curve.  Hence, metastable domains may survive up to about 10 T.  

\begin{figure}
\includegraphics[width=7.8cm]{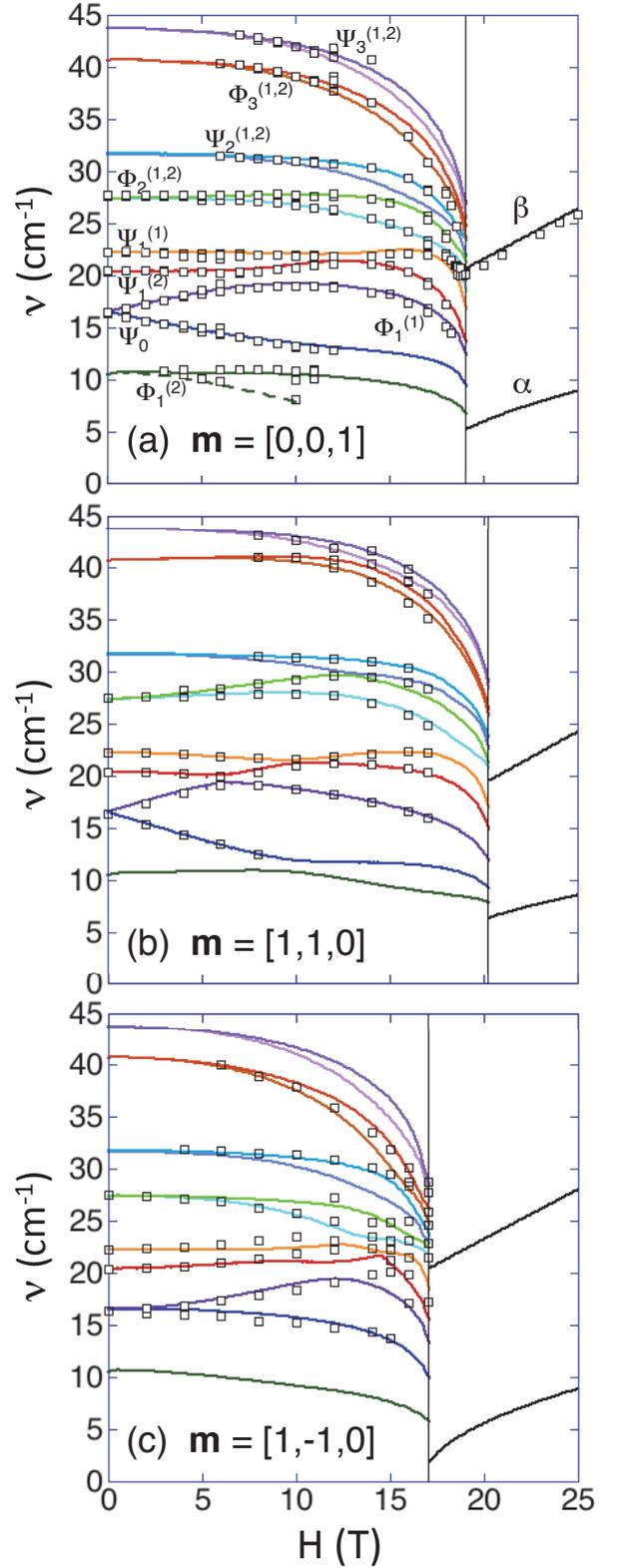}
\caption{(Color online) The theoretical mode spectra (solid curves) and experimental measurements (boxes) versus field for 
field orientations ${\bf m} =$ (a) $[0,0,1]$, (b) $[1,1,0]$, and (c) $[1,-1,0]$.  Solid vertical lines make the transition to the canted AF state.
The dashed curve in (a) indicates the predicted $\Phi_1^{(2)}$ for metastable domains 2 and 3.
}
\end{figure}

With $S_0 = 0.02$, the agreement between experiment and theory is even better than previously reported \cite{nagel13} for $\vm = [0,0,1]$
with $S_0=0.015$.  Nevertheless, that agreement deteriorates somewhat above 12 T, particularly for $\vm =[1,-1,0]$,
when avoided mode crossings strongly affect the mode frequencies.  It is possible that the trial spin state is not sophisticated enough at high magnetic fields.  
For example, the spin state in high magnetic fields may have a periodicity greater than two hexagonal layers.  

Above $H_c$, the canted AF state of Eq.(\ref{CAF}) supports only two modes that are labeled $\alpha $ and $\beta $ in Fig.4.
Because the transition at $H_c$ is first order, the spectroscopic modes change discontinuously at the critical field.

The estimates given above for $K$, $D_1$, and $D_2$ were based on fits to the 
four THz modes observed \cite{talbayev11, nagel13} in zero field \cite{exc}.  
Experimental data points in Fig.4 indicate that those four modes correspond to
$\Psi_0$/$\Phi_1^{(1)}$ (nearly degenerate), $\Psi_1^{(2)}$, $\Psi_1^{(1)}$, and $\Phi_2^{(1,2)}$ with frequencies 16.2, 20.7, 22.4, and 27.6 cm$^{-1}$, respectively.

\vfill

\section{Polarization matrix elements}

At zero magnetic field, only a few of the SW modes are optically active with finite magnetic-dipole resonance
matrix elements $\cvMa $, where
\begin{equation}
\vM =\frac{2\mb }{N} \sum_i \vS_i
\end{equation}
is the magnetization operator, $\vert 0 \rangle $ is 
the ground state with no SWs, and $\vert n \rangle $ is the $n$th excited state with a single SW mode at the 
cycloidal wavevector $\vQ $.   
At finite magnetic fields, all SW modes also have non-zero matrix elements $\cvPa $ of the induced electric 
polarization $\pin $.  The coexistence of the magnetic-dipole and polarization matrix elements is responsible for the NDD observed in the THz absorption spectra
for field along $[1,-1,0]$.  The physical mechanisms that contribute to $\pin $ below $\TN $
can be divided into three classes:  SC, MS, and ANI.  

For the SC- and MS-induced polarizations, we use LSDA+$U$ calculations \cite{junun} to simplify the matrices connecting the induced polarizations with the spin
operators.  This greatly reduces the number of polarization parameters.  In some instances, those matrices were simplified even further, 
either because some matrix elements were roughly equal or because 
additional matrix elements had a negligible effect on the NDD.  Those additional simplifications are described in Ref.[\onlinecite{junun}].
This section expresses the induced polarizations in the cycloidal reference frame $\{ \xq ,\yq ,\zq \}$.
In the laboratory reference frame $\{ x,y,z\}$, the induced polarizations are given in Appendices B, C, and D.

\subsection{SC-induced polarizations}

The SC-induced polarization $\psc $ is
produced by shifts in the O locations due to the hopping of electrons between 
Fe 3$d$ and O 2$p$ orbitals \cite{katsura05, jia07, kaplan11}.  The first SC-induced polarization is created by the well-known
inverse DM interaction \cite{katsura05, mostovoy06, sergienko06} corresponding to the $D_1$ term in the Hamiltonian.
This polarization can be generally written as
\begin{equation}
\label{SCM}
P^{{\rm SC}}_{1\alpha } = \sum_{\beta } \lambda^{(1)}_{\alpha \beta } T_{1 \beta },
\end{equation}
where $\vU_1 $ was defined by Eqs.(\ref{NDDT1}) and (\ref{NDDT3}).
According to Eq.(\ref{ULU}), the four nonzero matrix elements of $\underline{\lambda}^{(1)}$ are
$\lambda^{(1)}_{\xq \xq } = -\lambda^{(1)}_{\yq \yq } = \pm (c-d)$, $\lambda^{(1)}_{\yq \zq } = -2\sqrt{2}c$,
and  $\lambda^{(1)}_{\zq \yq } = -\sqrt{2}d$, where the plus sign is for domain 2 and the minus sign is for domains 1 or 3.

In a simplified version of the first SC-induced polarization with $c=d$, 
the diagonal terms $\lambda^{(1)}_{\xq \xq }$ and $\lambda^{(1)}_{\yq \yq }$
would vanish.  Then $P^{{\rm SC}}_{1\zq } = \lambda^{(1)}_{\zq \yq } T_{1 \yq }$
and $P^{{\rm SC}}_{1\yq } =\lambda^{(1)}_{\yq \zq } T_{1 \zq }$ would
reduce to the usual form \cite{katsura05} for the inverse DM interaction:
\begin{equation}
\label{scond}
P^{{\rm SC}}_{1\alpha } = -\sqrt{2} \bar{\lambda}^{(1)}_{\alpha } \Bigl\{ \xp \times \vU_1  \Bigr\}_{\alpha }, 
\end{equation}
with $\bar{\lambda}^{(1)}_{\xq }=0$, $\bar{\lambda}^{(1)}_{\yq } = \lambda^{(1)}_{\yq \zq}/\sqrt{2} $, and $\bar{\lambda}^{(1)}_{\zq } = -\lambda^{(1)}_{\zq \yq }/\sqrt{2} 
= -\lambda^{(1)}_{\yq \zq}/(2\sqrt{2})$ so that $\bar{\lambda}^{(1)}_{\zq }= -\bar{\lambda}^{(1)}_{\yq }/2$.

The second SC-induced polarization is associated with the DM interaction $D_2$:
\begin{equation}
P^{{\rm SC}}_{2\alpha }=  \lambda^{(2)}_{\alpha } T_{2 \alpha },
\end{equation}
where $\vU_2$ was defined by Eq.(\ref{NDDT4}).
As shown in Appendix B, the $\zq $ coefficient $\lambda^{(2)}_{\zq }$ may differ from the $\xq $ and $\yq $ coefficients $\lambda^{(2)}_{\xq } = \lambda^{(2)}_{\yq }$.

For the simple tilted cycloid of Eqs.(\ref{syc1}-\ref{syc3}) in zero magnetic field,
\begin{eqnarray}
\label{PSC1}
&&\langle \psc_1 \rangle =  2\pi S^2 \delta \cos \tau \nonumber \\
&&\biggl\{ \frac{1}{\sqrt{2}} \Bigl( \lambda^{(1)}_{\zq \yq }-\frac{1}{2}\lambda^{(1)}_{\yq \zq} \Bigr) \yp - \lambda^{(1)}_{\zq  \yq} \zp \biggr\} ,
\end{eqnarray}
\begin{equation}
\label{PSC2}
\langle \psc_2 \rangle = -\frac{3}{2}\lambda^{(2)}_{\zq } S^2 \sin 2\tau \, \zp .
\end{equation}  
If the cycloid were not tilted, only the first SC polarization would be nonzero.  When $c=d$, $\lambda^{(1)}_{\zq \yq }= \lambda^{(1)}_{\yq \zq}/2$
and the first term in $\langle \vP^{{\rm SC}}_1\rangle $
along $\yp $ vanishes.

\subsection{MS-induced polarizations}

The first MS-induced polarization is produced by the uniform displacement of Fe with respect to O:
\begin{equation}
P^{\rm MS}_{1\yq } = C_{1\yq } \, \yp \cdot \vW_1 ,
\end{equation}
\begin{equation}
P^{\rm MS}_{1\zq } = C_{1\zq } \, \zp \cdot \vW_1 ,
\end{equation}
\begin{equation}
W_{1u}=\frac{1}{N} \sum_{\langle i,j\rangle^{\bf u}} \,\vS_i \cdot \vS_j .
\end{equation}
It is easy to show that $\xp \cdot \vW_1 =0$.  For a simple twisted cycloid,
\begin{equation}
\langle \pms_1 \rangle = -\sqrt{3} C_{1\zq } S^2 \cos^2\tau \, \zp .
\end{equation}
The energy $-\vE \cdot \pms_1$ uniformly shifts all the nearest-neighbor interactions by
$\Delta J_1 = C_{1\zq } E_{\zq } /\sqrt{3}$.

The second MS-induced polarization can be written \cite{deSousa13}
\begin{equation}
\label{PMS2}
\pms_2= C_2\, \zp \times \vW_2 ,
\end{equation}
\begin{equation}
W_{2u}=  \frac{1}{N} \sum _{\langle i,j\rangle^{\vu }}  (-1)^\ni \,\vS_i \cdot \vS_j .
\end{equation}
Unlike $\vW_1$, $\vW_2$ alternates sign from one hexagonal layer to the next.
The cross product with $\zp $ in Eq.(\ref{PMS2}) ensures that $\pms_2$ remains a polar vector \cite{polar}.
For a simple tilted cycloid in zero field, $\langle \vW_2  \rangle = \langle \pms_2 \rangle =0$.  
The energy $-\vE \cdot \pms_2 $ shifts the 
nearest-neighbor exchange interaction $J_1$ by an amount proportional to $C_2 E$.  For example, the nearest-neighbor exchange
between spins at $\vR_i$ and $\vR_i + a\zz $ is shifted by $\Delta J_1 = (-1)^\ni C_2 (E_x-E_y)/\sqrt{3}$.   
Appendix C shows that $C_2= \sqrt{3}C_{1\yq }$.

The MS-induced polarization associated with next-nearest neighbor sites can be similarly constructed starting with
\begin{equation}
W_{3u}=\frac{1}{2N} \sum_{( i,j )'^{\bf u}} \,\vS_i \cdot \vS_j ,
\end{equation}
\begin{equation}
W_{4u}=\frac{1}{2N} \sum_{( i,j )'^{\bf u}} \,(-1)^{n_i}\, \vS_i \cdot \vS_j ,
\end{equation}
where all next-nearest neighbor pairs $(i,j)'^{\bf u}$ are double counted with $\vR_j -\vR_i = a\vv $, $\vert \vv \vert =\sqrt{2}$,
and $\vv \cdot \vu =0$.  So for $\vu =\vx $, $\vv = (0,1,1)$, $(0,1,-1)$, $(0,-1,1)$, and $(0,-1,-1)$.  For next nearest neighbors,
both $\vR_i$ and $\vR_j$ lie on either even or odd layers.

Since $\xp \cdot \vW_3  = \xp \cdot \vW_4 =0$, the polarizations associated with $\vW_3$ and $\vW_4$ are
\begin{equation}
P^{\rm MS}_{3,4\, \yq } = C_{3,4\, \yq } \, \yp \cdot \vW_{3,4} ,
\end{equation}
\begin{equation}
P^{\rm MS}_{3,4\, \zq } = C_{3,4\, \zq } \, \zp \cdot \vW_{3,4} .
\end{equation}
For a simple twisted cycloid, 
\begin{equation}
\langle {\bf P}^{\rm MS}_3\rangle = 2\sqrt{3} S^2 C_{3\zq }\, \zp 
\end{equation} 
while $\langle {\bf P}^{\rm MS}_4\rangle =0$.  
The energy $-\vE \cdot \pms_3$ uniformly shifts all the next nearest-neighbor interactions by
$\Delta J_2 = C_{3\zq } E_{\zq } /\sqrt{3}$.

Another possible MS-induced polarization is associated with the spin exchange ANI or different exchange couplings 
for different spin components $S_{i\alpha }$.
Because it is of order $\delta^2$, this polarization can be neglected.

\subsection{ANI-induced polarizations}

The ANI-induced polarization $\vP^{{\rm ANI}}=\vP^{{\rm ANI}}_{\perp } + \vP^{{\rm ANI}}_{\parallel }$, which arises from 
the spin-dependent hybridization between the Fe ions and their ligands, 
contains components perpendicular or parallel to $\zp$.
As shown in Appendix D, the perpendicular polarization 
 $\vP^{{\rm ANI }}_{\perp }=\vP^{{\rm ANI (1)}}_{\perp }+\vP^{{\rm ANI (2)}}_{\perp }$
has two sets of terms associated with the electric-field
dependence of the  local single-ion ANI axis  
$[\sin \theta_i \cos \phi_i ,\sin \theta_i \sin \phi_i, \cos \theta_i]$ defined by Eq.(\ref{LSIA}).  The first set is 
produced by the dependence of the polar angle $\theta_i $ on the electric field $\vE $:
\begin{equation}
\label{PAN1}
\vP^{{\rm ANI (1)}}_{\perp } =  \frac{ \xi_1  }{4N} \sum_i  \bigl(S_{i\xq }\xp  +S_{i\yq }\yp \bigr) S_{i\zq },
\end{equation} 
which agrees with the first ANI-induced polarization proposed by deSousa {\em et al.}
[\onlinecite{deSousa13}].    

An additional perpendicular polarization 
\begin{equation}
\frac{\xi_2}{4N}\sum_i 
\Bigl\{ \bigl({S_{i\xq }}^2 - {S_{i\yq }}^2 \bigr) \yp + 2 S_{i\xq }S_{i\yq } \xp \Bigr\}
\end{equation} 
with $\xi_2=\xi_1/(2\sqrt{2})$ was proposed in Ref.[\onlinecite{deSousa13}].
However, the cross terms $S_{i\beta }S_{i\gamma }$ ($\beta \ne \gamma $) in Eq.({\ref{PANL1}) 
cancel this contribution.

The second set of perpendicular ANI-induced terms is produced by the dependence of the azimuthal angle $\phi_i $ on $\vE $:
\begin{equation}
\label{PAN2}
\vP^{{\rm ANI (2)}}_{\perp } = -\frac{3\xi_3}{N} \sum_i  (-1)^{n_i} \bigl( S_{i\yq } \xp - S_{i\xq }\yp \bigr)S_{i \zq} ,
\end{equation}
which was not previously proposed.     

We also construct the ANI-induced polarization parallel to $\zp $ produced by the electric-field
dependence of the constant $K$:  
\begin{equation}
\vP^{{\rm ANI}}_{\parallel } = \frac{ \xi_4 }{4N}\, \zp \sum_i {S_{i\zq }}^2 ,
\end{equation}
which shifts the single-ion ANI by $\Delta K = \xi_4 E_{\zq }/4$.
For a simple tilted cycloid in zero field, 
$\langle \vP^{{\rm ANI}}\rangle = \xi_4 S^2 \zp /8$ includes only a contribution from $\xi_4 $ and 
is parallel to $\zp $.

\subsection{Total induced polarization}

With all proposed terms, the net induced polarization in the cycloidal phase is 
$\pin = \vP^{\rm SC} + \vP^{\rm MS} + \vP^{\rm ANI}$.
For the simple tilted cycloid,
\begin{eqnarray}
\label{pins}
&&\langle \zp \cdot \pin \rangle = S^2 \Bigl\{ -2\pi \lambda^{(1)}_{\zq \yq } \delta \cos \tau -\frac{3}{2} \lambda^{(2)}_{\zq } \sin 2\tau \nonumber \\
&&+ \frac{\xi_4 }{8} - \sqrt{3} C_{1\zq } \cos^2\tau +2\sqrt{3} C_{3\zq }\Bigr\}.
\end{eqnarray}
Of course, the components of $\langle \pin \rangle $ perpendicular to $\zp $ do not change the magnitude of the total polarization significantly.
The change in polarization from the paramagnetic phase above $\TN $ to the cycloidal phase
below $\TN $ is given by Eq.(\ref{pins}).  Recently, Lee {\em et al.}  [\onlinecite{lee13}] observed that
$\langle \pin \rangle $ has a magnitude of about 400 nC/cm$^2$ and opposes $\vP^{\rm FE}$ 
due to the suppressed displacement of the Fe ions compared to the Bi ions.  

By comparison, the induced polarization of the canted AF evaluated using Eq.(\ref{CAF}) is given by
\begin{eqnarray}
\label{PAF}
&&\langle \zp \cdot \paf \rangle = - S^2 \Bigl\{ 3\lambda_{\zq }^{(2)} \sin 2\tau +\sqrt{3} C_{1\zq } \cos 2\tau  \nonumber \\
&&+2\sqrt{3}C_{3\zq }\Bigr\},
\end{eqnarray}
which has no ANI contribution because the spins are in the $\{ \xq, \yq \}$ plane.  
So the change in polarization from the AF phase to the cycloidal phase at zero field is given by
\begin{eqnarray}
\label{CPI}
&&\Delta \langle \zp \cdot \pin \rangle = S^2 \Bigl\{ -2\pi \lambda^{(1)}_{\zq \yq } \delta \cos \tau \nonumber \\
&&+\frac{3}{2} \lambda^{(2)}_{\zq } \sin 2\tau +\frac{\xi_4}{8} -\sqrt{3} C_{1\zq } \sin^2 \tau \Bigr\}.
\end{eqnarray}
Despite an early measurement of 1 nC/cm$^2$ [\onlinecite{kadomtseva04}],
the magnitude of the polarization change $\Delta \langle \zp \cdot \pin \rangle $ below $H_c$ extrapolated to zero field
has recently been estimated as 40 nC/cm$^2$ [\onlinecite{tokunaga10, park11}].  

The Hamiltonian in zero electric field can be simply written in terms of the induced polarizations as
\begin{eqnarray}
\label{HP}
&&\frac{1}{N}{{\cal H}} = -\frac{\sqrt{3} J_1}{C_{1\zq }} \zp \cdot \vP^{\rm MS}_1  -\frac{\sqrt{3} J_2}{C_{3\zq }} \zp \cdot \vP^{\rm MS}_3
 \nonumber \\
 &&+\frac{\sqrt{2}D_1}{\lambda^{(1)}_{\zq \yq } }\, \zp \cdot \psc_1
+\frac{D_2}{\lambda^{(2)}_{\zq} }\, \zp \cdot \psc_2 
- \frac{4K}{\xi_4 } \, \zp \cdot \vP^{\rm ANI}\nonumber \\
&&- \frac{2\mb H}{N} \sum_i \vm \cdot \vS_i .
\label{Has}
\end{eqnarray}
Introducing the field dependence of the DM interactions, we find $\lambda^{(1)}_{\zq \yq } = -\sqrt{2}\partial D_1/\partial E_{\zq }$ and $\lambda^{(2)}_{\zq } = -\partial D_2/\partial E_{\zq}$.
Similarly, $C_{1\zq }=\sqrt{3} \,\partial J_1/\partial E_{\zq }$, $C_{3\zq }=\sqrt{3} \,\partial J_2/\partial E_{\zq }$, and $\xi_4= 4\partial K/\partial E_{\zq}$.  

All $\zp $ components of the induced polarization $\pin $
appear in ${\cal H}$ above.   
Because $\vP^{{\rm FE}}$ appears above $\TN $ in the paramagnetic phase, 
each static magnetically-induced polarization along $\zp $ corresponds to a 
term in the Hamiltonian. 
Due to the symmetry lowering associated with $\vP^{{\rm FE}}$, each bilinear spin term that 
appears in $\cal{H}$ also contributes to an induced polarization parallel to $\vP^{{\rm FE}}$.

Taking $c=d$ in Eq.(\ref{PSC1}), 
\begin{equation}
\langle \pin \rangle = -\frac{1}{N} \frac{\partial \langle {\cal H} \rangle }{\partial \vE }
\end{equation}
has no components perpendicular to $\zp $.
Components of the operator $\pin $ perpendicular to $\zp $ would then
contribute only to the transition matrix elements $\langle n \ne 0 \vert \pin \vert 0\rangle $.
In other words, ${\cal H}$ includes all induced polarizations with static contributions $\langle 0 \vert \pin \vert 0\rangle $
but not induced polarizations with only dynamical contributions $\langle n \ne 0\vert \pin \vert 0\rangle $.   For example,
$\vP^{{\rm MS}}_2$ does not appear in $\cal{H}$ because $\langle \vP^{{\rm MS}}_2\rangle =0$.

We used Eq.(\ref{HP}) to check our numerical results for the matrix elements $\langle n \vert \zp \cdot \vP  \vert 0\rangle $.  Since 
$\langle n \vert {\cal{H}} \vert 0 \rangle = E_0 \delta_{n0}$,
the appropriate sum of polarization matrix elements with the field-dependent term $-N H \langle n \vert \vm \cdot \vM \vert 0 \rangle $ 
must vanish when $n \ne 0$.  We verified that this condition is indeed satisfied for all excited states and magnetic fields.

\section{THz Absorption}

The absorption of THz light is given by $\alpha (\omega ) = (2\omega /c)\,{\rm{Im}} N(\omega )$
where \cite{miyahara11, miyahara14}
\begin{equation}
N(\omega ) \approx \sqrt{(\ein_{ii} + \chi^{ee}_{ii} (\omega )) (1+\chi^{mm}_{jj}(\omega ))}\pm  \chi^{me}_{ji}(\omega )
\end{equation}
is the complex refractive index for a linearly polarized beam, $\underline{\chi}^{ee}$, $\underline{\chi}^{mm}$ and $\underline{\chi}^{me}$ are the dielectric, magnetic, and 
magnetoelectric susceptibility tensors describing the dynamical response of the spin system \cite{kezsmarki11, bordacs12, miyahara11, kezsmarki14} and 
$\underline{\epsilon }^{\infty }$ is the background dielectric constant tensor associated with charge excitations at higher energies.  
Subscripts $i$ and $j$ refer to the electric and magnetic polarization directions, respectively.
The second term, which depends on the light propagation direction
and produces NDD, is separated from the mean absorption
by writing $N(\omega ) =\bar{N}(\omega )\pm \chi^{me}_{ji} (\omega )$.

Summing over the SW modes $n$ at the cycloidal ordering wavevector $\vQ $,
$\Delta \alpha (\omega ) = (4\omega /c)\, {\rm{Im}} \chi^{me}_{ji} (\omega ) $ is given by 
\begin{equation}
\label{dal}
\Delta \alpha (\omega ) = \sum_n A_n \, \delta (\omega - \omega_n),
\end{equation}
\begin{equation}
A_n= N X \omega_n \, \rm{Re } \Bigl\{ \rho_{n0} \mu_{0n} \Bigr\} ,
\end{equation}
\begin{equation} 
\rho_{0n} = \langle 0\vert \pin \cdot \ve /\cV \vert {\it n} \rangle ,
\end{equation}
\begin{equation}
\mu_{0n}= \langle 0\vert \vM \cdot \vh /\mB \vert {\it n} \rangle ,
\end{equation}
where $\cV=a^3$ is the volume per Fe site, $\pin /\cV $ is given in units of nC/cm$^2$ and 
\begin{equation}
X= \frac{4\pi \mb }{\hbar } \frac{{\rm nC}}{{\rm cm}^2}= \frac{0.1388}{{\rm cm}}.
\end{equation}
The THz electric and magnetic fields are polarized in the $\ve $ and $\vh $ directions, respectively.  

After expanding $\bar{N}(\omega )$
for small susceptibilities, we find that $\mal (\omega ) =(2\omega /c)\, {{\rm{Im}}} {\bar N}(\omega )$
is given by
\begin{equation}
\mal (\omega ) = \sum_n B_n\, \delta (\omega - \omega_n),
\end{equation}
\begin{equation}
B_n = N \omega_n \Bigl\{ Y_1  \vert \rho_{0n} \vert^2 + Y_2 \vert \mu_{0n }\vert^2 \Bigr\},
\end{equation}
where 
\begin{equation}
Y_1 = \frac{\pi \cV }{ \hbar c\epsilon_0 \sqrt{\ein }} \frac{{\rm nC}^2}{{\rm cm}^4} = \frac{6.975 \times 10^{-4}}{\sqrt{\ein }\, {\rm cm }},
\end{equation}
\begin{equation}
Y_2 = \frac{\pi \mb^2 \mu_0 \sqrt{\ein }}{\hbar c \cV } = \frac{1.727 \sqrt{\ein }}{{\rm cm}}.
\end{equation}
Notice that $X = 4\sqrt{Y_1 Y_2}$.

The dielectric constant $\ein_{ii}$ depends on the polarization $\ve $ of light. 
Based on a fit to the interference fringes, $\ein = 27.54$ and 51.55 for $\ve =  [1,1,0]$ and $\ve = [1,-1,0]$, respectively.

For each orientation of the static magnetic field and light polarization, 
the integrated weight of every spectroscopic peak at $\omega_n$ is compared with the measured values.   
This eliminates estimates of the individual peak widths.  Because the polarization and magnetization matrix elements 
are generally complex with an arbitrary overall phase that differs for each mode $n$, we can choose
$\langle 0 \vert P_{\zq } \vert n \rangle $ to be real.  Other magnetization and polarization 
matrix elements for mode $n$ are then either purely real or imaginary.
Under reversal of the field orientation, our numerical results indicate that 
$\langle 0 \vert \vM \vert n\rangle \rightarrow -\langle 0 \vert \vM \vert n \rangle^{\star }$
and $\langle n \vert \vP \vert 0\rangle \rightarrow \langle n \vert \vP \vert 0 \rangle^{\star }$. 
It follows that the NDD vanishes for zero field.
Our numerical results also indicate that the NDD should vanish \cite{symv} 
for field directions $[0,0,1]$ and $[1,1,0]$.

\begin{table}
\caption{\textbf{Fitting parameters (nC/cm$^2$) from $\Delta \alpha $}}
\begin{ruledtabular}
\begin{tabular}{lllllllc}
 & $ \lambda^{(1)}_{\yq \zq } $ & $\lambda^{(1)}_{\zq \yq }$ & $\lambda^{(2)}_{\xq }$, $\lambda^{(2)}_{\yq }$ & $\lambda^{(2)}_{\zq }$  & $N_{{\rm par}}$ & $\chi^2_{{\rm min}}$  \\
 \hline 
fit 1  & $-82.0$ & $-50.3$ & $+35.2$ & $+13.6$ & $\,\,\, 4$ &1.543  &  \\
 error  &$\pm 3.1$ & $\pm 8.0$  & $\pm 1.9$ & $\pm 3.0$  &  & \\
 \hline 
 fit 2  & $-78.7$ & $-39.4$ & $+33.7$ & $+13.9$ & $\,\,\, 3$ & 1.536 & \\
 error  &$\pm 3.5$ & $\pm 1.7$  & $\pm 1.9$ & $\pm 3.0$  &  &  \\
\end{tabular}
\end{ruledtabular}
\end{table}

Unfortunately, fitting results for the mean absorption $\mal (\omega )$ were markedly inferior to results for $\Delta \alpha (\omega )$.  
This may be caused by uncertainty about the dielectric constants $\ein_{ii}$, which does not enter
$\Delta \alpha (\omega )$.  Moreover, the measured {\it difference} between the absorption in positive and negative fields
is much less prone to systematic experimental error than $\mal (\omega )$.

Experimental results for the NDD with field along $\vm = [1,-1,0]$ are plotted in Figs.5(a) and (b) for $\ve = [1,-1,0]$ and $[1,1,0]$, respectively.
For some modes, the NDD is strong enough that $\alpha (\omega )$ is small for light traveling in one direction but
large for light traveling in the other direction \cite{istun}.  In particular, for $\Psi_0\approx 15.5$ cm$^{-1}$ and $\ve =[1,1,0]$,
$A_n = 0.67$ cm$^{-2}$ for an 8 T field along $\vm = [1,-1,0]$ and light propagating along $\vk =[0,0,1]$ while
$A_n = 4.12$ cm$^{-2}$ when either $\vm $ or $\vk $ is reversed.

Fits to the NDD are based on the plotted 2, 4, 6, 8, 10, and 12 T data sets.
For each field value with two light polarizations, we evaluate the integrated weights for the 8 modes 
$\Psi_0$, $\Phi_1^{(1)}$, $\Psi_1^{(1,2)}$, $\Phi_2^{(1,2)}$, and $\Psi_2^{(1,2)}$ between roughly 12 and 35 cm$^{-1}$.
Hence, there are $N_{{\rm dat}}=96$ data points for $\Delta \alpha (\omega )$.  Because the $\Phi_2^{(1,2)}$
and $\Psi_0$/$\Phi_1^{(1)}$ pairs are nearly degenerate in small fields, the average predicted NDD of those 
pairs are compared with the measured values of $\Delta \alpha (\omega )$ in a 2 T field.

Remarkably, the NDD for $\vm = [1,-1,0]$
is dominated by the two sets of SC polarizations $\vP^{{\rm SC}}_1$ and $\vP^{{\rm SC}}_2$ associated
with the DM interactions $D_1$ and $D_2$, respectively.  Neglecting MS and ANI
leaves four polarization parameters:  $\lambda^{(1)}_{\yq \zq }$, $\lambda^{(1)}_{\zq \yq}$, $\lambda^{(2)}_{\xq }
=\lambda^{(2)}_{\yq }$, and $\lambda^{(2)}_{\zq}$.
The diagonal parameters $\lambda^{(1)}_{\xq \xq }= - \lambda^{(1)}_{\yq \yq } = c-d$ (domain 2) are obtained from
$\lambda^{(1)}_{\yq \zq }=-2\sqrt{2}c$ and $\lambda^{(1)}_{\zq \yq }=-\sqrt{2} d$ using the relations in Appendix B.  

Above about 12 T, agreement between the theoretical and experimental values for $\Delta \alpha $ begins to deteriorate.  This 
failure may be caused by the avoided mode crossings between 12 and 16 T, which mix
$\Psi_1^{(2)}$, $\Psi_1^{(1)}$, and $\Phi_2^{(1)}$ and are imperfectly captured by our model.
Since each of those modes exhibits pronounced NDD, the fitted polarization parameters are very sensitive to the precise behavior of the modes
at the avoided crossings.

Based on the typical noise level for the absorption, we 
set the experimental uncertainty for the integrated NDD of each peak to $\sigma = 1$ cm$^{-2}$.  
The error bars for each polarization parameter are then 
obtained from the condition that $\chi^2/\chi^2_{{\rm min}}$ increases by \cite{stat} $1/(N_{{\rm dat}}-N_{{\rm par}}-1)$.  

The results for fit 1 in Table II with $N_{\rm par} =4$ free parameters indicate that for domain 2, $\lambda^{(1)}_{\xq \xq } =
-\lambda^{(1)}_{\yq \yq } \approx -6.6 \pm 6.7$ nC/cm$^2$. 
So our results imply the absence of the diagonal terms in $\underline{\lambda }^{(1)}$
and support the simplified form of Eq.(\ref{scond}) for the first SC-induced polarization.  With a slightly smaller $\chi^2_{{\rm min}}$,
fit 2 with $N_{{\rm par}} =3$ takes $\lambda^{(1)}_{\xq \xq }=\lambda^{(1)}_{\yq \yq }=0$ and 
$\lambda^{(1)}_{\yq \zq }=2\lambda^{(1)}_{\zq \yq}$.  

\begin{figure}
\includegraphics[width=8.8cm]{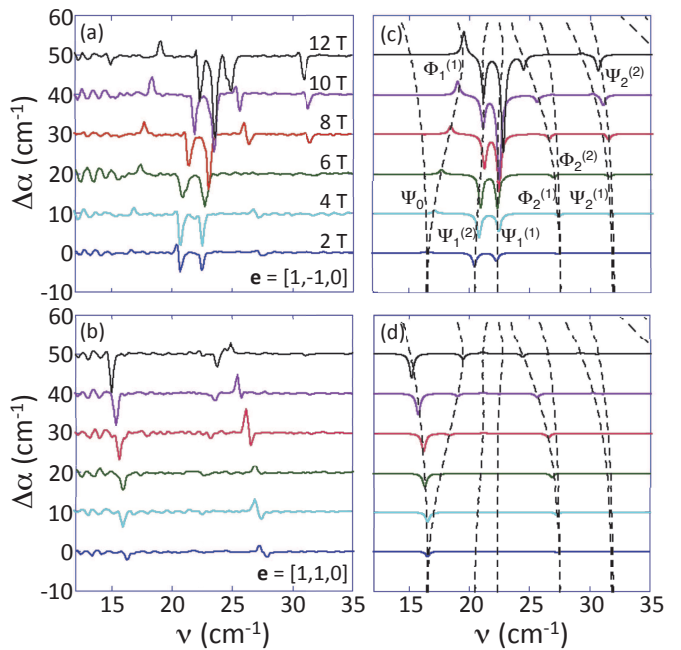}
\caption{(Color online) The measured (a,b) and predicted (c,d) NDD for 2 to 12 T fields along $\vm = [1,-1,0]$ and for $\ve = [1,-1,0]$ (a and c) or $[1,1,0]$ (b and d).
Predictions are based on fit 2.  Dashed curves in (c) and (d) are the predicted mode frequencies.
}
\end{figure}

Because the sample $\vP^{{\rm FE}}$ may point parallel or antiparallel to [1,1,1], the overall sign
of $\Delta \alpha (\omega )$ and of the polarization parameters is ambiguous.
According to Katsura {\em et al.} \cite{katsura05}, however, $\bar{\lambda}_{\zq } = -\lambda^{(1)}_{\zq \yq }/\sqrt{2}$ in Eq.(\ref{scond}) 
should be positive so that $\lambda^{(1)}_{\zq \yq }<0$.  
This condition is used to fix the overall sign of the SC parameters in Table II.

Results for fit 2 are plotted in Figs.5(c) and (d).  Although it underestimates the NDD for $\Phi_2^{(1)}$
when $\ve = [1,1,0]$ (perhaps due to a small shift in the spectra $\alpha (\omega )$ for positive and negative fields), 
this fit otherwise describes all the relevant features of the NDD with field orientation $\vm = [1,-1,0]$.
Based on fit 2 parameters, the predicted SC polarizations $\langle \vP^{{\rm SC}}_1\rangle  \approx 7.0 \,\zp $ nC/cm$^2$ and 
$\langle \vP^{{\rm SC}}_2 \rangle \approx -2.1\, \zp$ nC/cm$^2$ point parallel and antiparallel to $\vP^{{\rm FE}}$, respectively.  
Hence, the total SC-induced polarization points along $\vP^{{\rm FE}}$ with a magnitude of 4.9 nC/cm$^2$.  
From Eq.(\ref{PAF}), the SC-induced polarization of the canted AF above $H_c$ is given by $-4.2 \,\zp $ nC/cm$^2$
opposite to $\vP^{{\rm FE}}$.

Since the MS-induced polarizations above and below $H_c$ differ by 
$-\sqrt{3} S^2C_{1\zq } \sin^2 \tau \approx 6.9 \times 10^{-4} \, C_{1\zq }$,
this term can be ignored in Eq.(\ref{CPI}).
Using the LSDA+$U$ result [\onlinecite{junun}]
$\xi_4 \approx 110$ nC/cm$^2$, the change in $\langle \zp \cdot \pin \rangle $ below $H_c$
contains the ANI-induced contribution $S^2\xi_4 /8 \approx 86$ nC/cm$^2$.
Therefore, the total predicted change $\Delta \langle \zp \cdot \pin \rangle \approx 96$ nC/cm$^2$ in the 
induced polarization from above to below $H_c$
is more than twice larger than the recent experimental estimates of 40 nC/cm$^2$ [\onlinecite{tokunaga10, park11}].
Alternatively, fitting the experimental jump to Eq.(\ref{CPI}) gives $\xi_4 = 40$ nC/cm$^2$, 40\% smaller than the LSDA+$U$ prediction.

A possible explanation for this discrepancy is that we have slightly overestimated $D_2$ and $S_0$.  As mentioned above, 
taking $D_2 < 0.079$ meV or $S_0 < 0.185$ would stabilize a canted AF phase with spins tilted out of the 
$\{ \xq ,\yq \}$ plane.  Because such a state would recoup some ANI energy, the predicted jump in the induced polarization 
at $H_c$ would be reduced.  The planar canted AF phase of Eq.(\ref{CAF}) would then appear
above some higher critical field $H_c' > H_c$.  Due to the non-coplanar AF phase, 
the lower AF mode $\alpha $ in Fig.4 would decrease with field between 
$H_c$ and $H_c'$, vanish at $H_c'$, and increase with field only above $H_c'$.

\begin{figure}
\includegraphics[width=8.8cm]{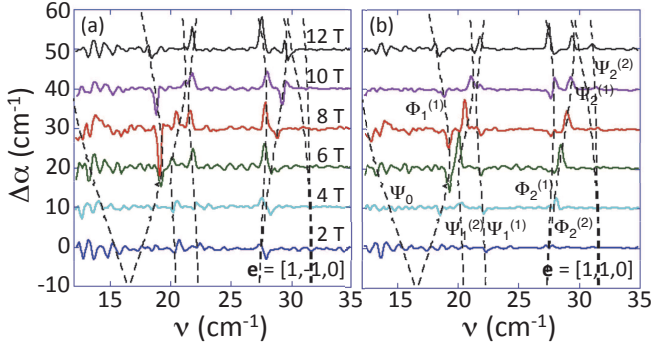}
\caption{(Color online) The measured NDD for 2 to 12 T fields along $\vm = [1,1,0]$ and for $\ve =$ (a) $[1,-1,0]$ or (b) $[1,1,0]$.
Dashed curves are the predicted mode frequencies.
}
\end{figure}

Using
$\lambda^{(1)}_{\zq \yq } = -\sqrt{2}\partial D_1/\partial E_{\zq }$ and $\lambda^{(2)}_{\zq } = -\partial D_2/\partial E_{\zq}$,
the results of fit 2 for $\lambda^{(1)}_{\zq \yq }$ and $\lambda^{(2)}_{\zq }$ can be used to evaluate the dependence of $D_1$ and $D_2$ on an 
electric field applied along $\zp $.   Raman measurements  \cite{rov10} indicate that the spectroscopic modes exhibit significant dependence on an electric field
of 75 kV/cm along [0,1,0].  For an electric field of 100 kV/cm along [1,1,1], we find $\Delta D_1/D_1=6.0 \times 10^{-3}$ and $\Delta D_2/D_2= -6.4\times 10^{-3}$.
Although very small, the change in $D_1$ will slightly increase the size of $\delta $ and reduce the period of the cycloid.
The change in $D_2$ will slightly reduce the tilt angle $\tau $.

\section{Discussion}

Although the distorted cycloid of \BF is produced by the competition between magnetic interactions,
the SC polarization dominates the NDD of \BP .   The NDD of \BF along $\vm =[1,-1,0]$ is well described by our model.   
Due to the pronounced NDD for $\Psi_0$ when $\ve =[1,1,0]$,  \BF may be used as an optical diode, transparent to light 
traveling in one direction but opaque for light traveling in the opposite direction.  Despite the successes of this model, 
several issues must be addressed.  

For light propagating along $\vk = [0,0,\pm 1]$, symmetry arguments \cite{symv} and our numerical results indicate that
NDD should be absent for $\vm = [\eta, \eta ,\kappa ]$ with stable domain 1 if either $\ve $ or $\vh $ coincides with $\xp_1 $.  
Even for $\vm =[1,1,1]$, where all three domains 
are degenerate \cite{fishman13c}, the NDD should
vanish if domains 2 and 3 are equally populated.  While NDD is not observed for $\vm = [0,0,1]$, 
the NDD for $\vm = [1,1,0]$ is plotted in Fig.6.  The most pronounced NDD is seen near the avoided mode crossing of
$\Psi_1^{(2)}$ and $\Phi_1^{(1)}$ around 8 T.  

To estimate the relative sizes of the NDD in fields along $[1,1,0]$ and $[1,-1,0]$, we calculate the net squared NDD,
\begin{equation}
\Ups = \frac{1}{N_{\rm dat}\, \sigma^2 } \sum_n {A_n}^2 ,
\end{equation}
where the sum runs over all modes between 2 and 12 T and $A_n$ was defined by Eq.(\ref{dal}).
Since $\Ups = 3.50$ and 9.45, respectively, the observed NDD is substantially weaker for $[1,1,0]$ than for $[1,-1,0]$.
Because the contributions from metastable domains 2 and 3 cancel each other, they 
can not explain the NDD observed for $\vm = [1,1,0]$.  While a population imbalance between metastable domains 2 and 3 would produce
very weak NDD, domain 1 is expected to predominate above a few Tesla.
Misalignment of the crystal 
could produce the observed NDD when $\vm = [1,1,0]$ but the excellent agreement between the measured and predicted mode
spectrum in Fig.4(b) suggests that the sample is aligned quite well. 

The NDD for $\vm =[1,1,0]$ probably arises from
an optical misalignment \cite{mis} with the polarization vectors $\ve $ and $\vh $ rotated about
$\vk = [0,0,1]$.  For $\vm = [\eta ,\eta ,\kappa ]$, $\Ups (\alpha ) \approx \Ups (\pi/4) \sin^2 (2\alpha )$, 
where $\alpha $ is the angle between $\ve $ and $[1,-1,0]$.  As shown in Fig.7 for $\vm = [1,1,0]$ and $[0,0,1]$,
$\Ups (\alpha ) $ peaks at $\alpha = \pi /4$, i.e.
when $\ve =[1,0,0]$ and $\vh =[0,1,0]$ or $\ve = [0,1,0]$ and $\vh = [-1,0,0]$.  For $\vm = [1,-1,0]$, $\Ups (\alpha ) \approx \Ups (0 ) - (\Ups (0)/2) \sin^2(2\alpha )$ is
predicted to drop to about $\Ups (0)/2$ at $\alpha = \pi /4$.
Note that the squared NDD for an individual mode does not obey these relations:  they are obeyed only by the net squared NDD summed over all modes.
Measuring the NDD while rotating the THz polarization vectors about $[0,0,1]$ would help resolve questions about the NDD when $\vm =[1,1,0]$.

While the SC dominates the dynamical response of \BP , MS dominates its static properties.
As argued elsewhere \cite{jununp}, $\langle \vP^{{\rm MS}}_1 \rangle \approx -\sqrt{3} S^2 C_{1\zq }\zp $ 
dominates the induced polarization and opposes $\vP^{{\rm FE}}$ below $\TN $, in agreement with recent measurements \cite{lee13}.  
A unified model of ferroelectricity \cite{xiang13, lu15} also concludes that SC and ANI make minor contributions to $\langle \pin \rangle $ compared to MS.  

\begin{figure}
\includegraphics[width=8cm]{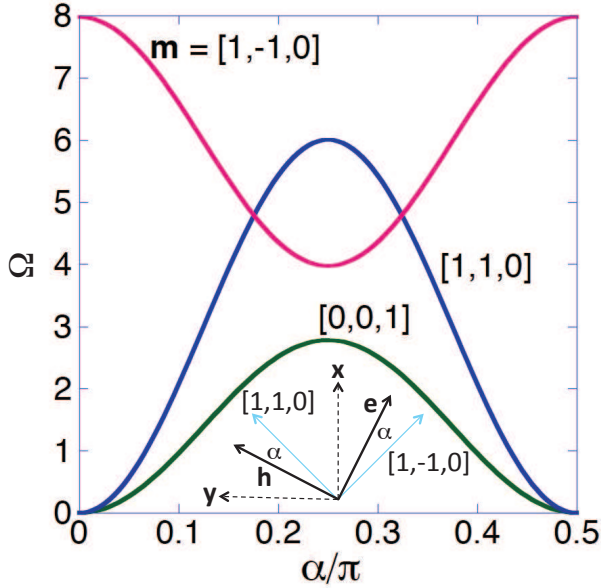}
\caption{(Color online) The predicted $\Ups $ versus $\alpha $ for fields
along $[1,-1,0]$ (red), $[1,1,0]$ (blue), and $[0,0,1]$ (green).  SC parameters are obtained from fit 2.  Inset shows the
rotation of $\ve $ and $\vh $ about $[0,0,1]$.
}
\end{figure}

The distinction between static and dynamics properties in \BF is not surprising.  Since spin fluctuations $\delta \vS_i$ 
are transverse to the almost collinear, cycloidal spin state $\langle \vS_i \rangle $, we find that $\delta \vS_i \times \langle \vS_j \rangle \ne 0$ but
$\delta \vS_i \cdot \langle \vS_j \rangle \approx 0$
for nearby sites $i$ and $j$.  Because the ANI is \BF is extremely weak,
spin fluctuations more strongly affect the SC-induced polarization than the MS- and ANI-induced polarizations.  
By contrast, the almost collinear spin structure of \BF efficiently produces a static polarization through the MS and ANI but not through the SC since  
$\langle \vS_i \rangle \cdot \langle \vS_j \rangle \neq 0$ and $(1/N) \sum_i \langle S_{i \zq } \rangle \langle S_{i \zq } \rangle \neq 0$
but $\langle \vS_i\rangle \times \langle \vS_j\rangle \approx 0$.
 
Tokunaga {\em et al.} \cite{tok15} recently attributed the induced {\it transverse} polarization along $\yp $ to the 
first SC polarization $\vP^{{\rm SC}}_1$ with $\underline{\lambda}^{(1)}$ matrix elements $\lambda^{(1)}_{\zq \yq }$ and $\lambda^{(1)}_{\yq \yq }$.  
Those authors found that $\vert \lambda^{(1)}_{\yq \yq }\vert \approx 104$ nC/cm$^2$ and $\vert \lambda^{(1)}_{\zq \yq }\vert \approx 73$
nC/cm$^2$.  By contrast, fit 1 indicates that $\vert \lambda^{(1)}_{\yq \yq } \vert \approx 6$ nC/cm$^2$ is very small.  
The result $\vert \lambda^{(1)}_{\zq \yq } \vert $ from Ref.[\onlinecite{tok15}] is reasonably close to the result $\vert \lambda^{(1)}_{\zq \yq }\vert \approx 50 \pm 8$ nC/cm$^2$
from fit 1.
  
Considering only the first set of SC terms associated with $D_1$, earlier work \cite{fishman13a} identified $\Psi_1^{(1)}$ as
an electromagnon \cite{miy12, chen13} that can be excited by a THz electric field when $\vH =0$. 
When both sets of SC terms are considered, $\Psi_0$/$\Phi_1^{(1)}$ and $\Phi_2^{(1,2)}$
also become electrically active at zero field.  Of the four modes observed in zero field, only $\Psi_1^{(2)}$
at 20.4 cm$^{-1}$ is not electrically active.  Using the SC parameters in Table II, $\Psi_1^{(1)}$ couples
most strongly of all modes to a THz electric field for domains 2 and 3.  
This mode also exhibits the strongest NDD for nonzero field.
 
To summarize, the SC polarization matrix elements dominate the NDD in \BP .  But work remains to 
understand the origin of the MS-induced and perpendicular polarizations in this important material.   
Our explanations for the jump in the induced polarization at $H_c$ 
and for the observed weak NDD when $\vm = [1,1,0]$ need to be confirmed.  
Nevertheless, we believe that the present work on \BF provides a compelling example of how a quantitative 
microscopic theory of magnetoelectric couplings follows from an analysis of the observed dynamical
magnetoelectric response based on an effective spin model supplemented by first-principles calculations.

We acknowledge helpful conversations with Eric Bousquet and Rogerio deSousa.   
We also thank Hee Taek Yi and Sang-Wook Cheong for preparation of the BiFeO$_3$ sample.
Research sponsored by the Department of Energy, Office of Sciences, Basic Energy Sciences, Materials Sciences and Engineering Division (RF and JL)
and by the Hungarian Research Funds OTKA K
108918, OTKA PD 111756, and Bolyai 00565/14/11(SB and IK).
TR and UN acknowledge support by the Estonian Ministry of Education and Research Grant
IUT23-03 and by the Estonian Science Foundation Grant ETF8703.

\appendix

\section{Experimental Details}

A  single ferroelectric domain BiFeO$_3$ sample with face area 7\,mm$^2$ $(0,0,1)$ and thickness $d=0.37$\,mm along $[0,0,1]$ was grown at
Rutgers University.  Voigt measurements ($\mathbf{k}\perp \mathbf{m}$) up to 17 T were performed in Tallinn.
Faraday ($\mathbf{k}\parallel \mathbf{m}$) measurements up to 12 T were performed in Tallinn and up to 31 T in Nijmegen, as reported earlier \cite{nagel13}.

The Tallinn laboratory uses a Martin-Puplett type interferometer with a Si bolometer operating at  0.3\,K and a mercury arc light source.
Light pipes direct light to the sample in a He exchange gas-filled sample chamber placed into the cold 52\,mm bore of a vertical-field superconducting 17 T solenoid.
In the Voigt configuration, mirrors before and after the sample change the light direction perpendicular to $\mathbf{m}$.
A rotatable wire grid on the dielectric substrate polarizer is placed before the first mirror. 
The sample can be rotated about the axis parallel to the direction of light propagation. 
A set of low pass filters with different cut-off frequencies is situated on the filter wheel in liquid He between the sample chamber and the bolometer chamber below the solenoid.

Applying a 17 T field at 4 K for tens of minutes populates magnetic domain 1 when $\mathbf{m} = [1,1,0]$ or 
domains 2 and 3 when $\mathbf{m} = [1,-1,0]$ [\onlinecite{nagel13}].
Spectra were then measured in different $\pm \vH $ fields for about 15 minutes per field.
No change in the magnetic domain populations was observed when a $-17$ T field was applied after a $+17$ T field.

The zero-field absorption spectrum was subtracted from the spectra measured in field, thereby
canceling out diffraction and interference effects caused by the sample.
The differential absorption coefficient is $\alpha(H)-\alpha (0)=-\ln(I_H/I_0)/d$,
where $I_0$ and $I_H$ are light intensity spectra in zero and $H$ field and $d$ is the sample thickness.
Negative peaks in the differential absorption spectra for all field values were used to calculate the zero-field spectrum.
To generate the field-dependent spectra, the calculated zero-field spectrum was aNDDed to the differential spectra.
The NDD spectra $\Delta \alpha = \alpha(\vH )-\alpha (-\vH )$ do not depend on the zero-field spectra.

\section{SC-induced polarizations} 

The $D_1$ term in the Hamiltonian can be written
\begin{equation}
\label{VSC1}
V^{{\rm SC}}_1 =\sum_{u, \langle i,j\rangle^{\bf u} } {\bf{F}}^{(u)} \cdot (\vS_i \times \vS_j ).
\end{equation}
In the absence of an electric field, 
\begin{equation}
{\bf F}^{(x)} = \frac{D_1 }{\sqrt{3}}(0,1,-1),
\end{equation}
\begin{equation}
{\bf F}^{(y)}  = \frac{D_1}{\sqrt{3}}( -1, 0, 1),
\end{equation}
\begin{equation}
{\bf F}^{(z)} = \frac{D_1}{\sqrt{3}}(1, -1,0)
\end{equation}
along $\vx $, $\vy $, and $\vz $, respectively.
The polarization associated with $D_1$ is then
\begin{eqnarray}
P^{{\rm SC}}_{1\alpha } && = -\frac{1}{N} \frac{\partial V_1^{\rm SC}}{\partial E_{\alpha }} \nonumber \\
&&= \frac{1}{N} \sum_{u, \langle i,j\rangle^{\bf u} ,\beta }  \Pi^{(u)}_{\alpha \beta }\, (\vS_i \times \vS_j )_{\beta },
\end{eqnarray}
where $\Pi^{(u)}_{\alpha \beta } = - \partial F^{(u)}_{\beta }/\partial E_{\alpha } $.  LSDA+$U$ calculations \cite{junun} reveal that
\begin{equation}
\underline{\Pi}^{(x)} =
\left(
\begin{array}{ccc}
0 & -d & d \\
0 & -c & -c \\
0 & c & c \\
\end{array} \right),
\end{equation}
\begin{equation}
\underline{\Pi}^{(y)} =
\left(
\begin{array}{ccc}
c & 0 & c \\
d & 0 & -d \\
-c & 0 & -c \\
\end{array} \right),
\end{equation}
\begin{equation}
\underline{\Pi}^{(z)} =
\left(
\begin{array}{ccc}
-c & -c & 0 \\
c & c & 0 \\
-d & d & 0 \\
\end{array} \right).
\end{equation}
Consequently, 
$\partial {\bf F}^{(u)}/\partial E_{\alpha }$ is not parallel to ${\bf F}^{(u)}$.

In the lab reference frame $\{ x, y, z \}$,
regrouping terms for domain 2 yields $P^{{\rm SC}}_{1\alpha } = \sum_{\beta } \Lambda^{(1)}_{\alpha \beta } T_{1 \beta }$ with 
$\underline{\Lambda }^{(1)}=\underline{\Pi }^{(x)}-\underline{\Pi}^{(z)}$ or
\begin{equation}
\underline{\Lambda }^{(1)} =
\left(
\begin{array}{ccc}
c & c-d & d \\
-c & -2c & -c \\
d & c-d &  c\\
\end{array} \right).
\end{equation}
We transform this matrix into the cycloidal reference frame $\{ \xq ,\yq ,\zq \}$
using the unitary matrix $\underline{U}$ for domain 2: 
\begin{equation}
\underline{U} =
\left(
\begin{array}{ccc}
1/\sqrt{2} & 0 & -1/\sqrt{2} \\
-1/\sqrt{6}  & \sqrt{2/3} & -1/\sqrt{6} \\
1/\sqrt{3} & 1/\sqrt{3} &  1/\sqrt{3} \\
\end{array} \right)
\end{equation}
so that
\begin{eqnarray}
\label{ULU}
&&\underline{\lambda}^{(1)} = \underline{U} \, \underline{\Lambda}^{(1)} \, \underline{U}^{-1} \nonumber \\
&& =
\left(
\begin{array}{ccc}
c-d & 0 & 0 \\
0 & d-c & -2\sqrt{2}c \\
0 & -\sqrt{2}d &  0\\
\end{array} \right).
\end{eqnarray}
The diagonal terms change sign for domains 1 and 3.

The $D_2$ term in the Hamiltonian can be written $V_2^{{\rm SC}} = D_2 N \, \zp \cdot \vU_2 $.
Thus, the SC-induced polarization associated with $D_2$ can be generally 
written $P^{{\rm SC}}_{2 \alpha } = \sum_{\beta }\Lambda^{(2)}_{\alpha \beta }\, T_{2 \beta }$.
In the lab reference frame, $\underline{\Lambda }^{(2)}$ is given by \cite{junun}
\begin{equation}
\underline{\Lambda}^{(2)} =
\left(
\begin{array}{ccc}
e & f & f \\
f & e & f \\
f & f &  e\\
\end{array} \right).
\end{equation}
Transforming into the cycloidal reference frame,
\begin{equation}
\underline{\lambda}^{(2)} = \underline{U} \, \underline{\Lambda }^{(2)} \, \underline{U}^{-1} =
\left(
\begin{array}{ccc}
e-f & 0 & 0 \\
0 & e-f & 0 \\
0 & 0 &  e+2f \\
\end{array} \right)
\end{equation}
for all three domains.
So $\underline{\lambda}^{(2)}$ is diagonal with components $\lambda_{\xq }^{(2)} = \lambda_{\yq }^{(2)}= e-f$ and 
$\lambda_{\zq }^{(2)}= e+2f$.

$\,$

\section{MS-induced polarizations} 

The MS-induced polarizations are $P^{{\rm MS}}_{1 \alpha } = \sum_{\beta }\Gamma^{(1)}_{\alpha \beta }\, W_{1 \beta }$ and
$P^{{\rm MS}}_{2 \alpha } = \sum_{\beta }\Gamma^{(2)}_{\alpha \beta }\, W_{2\beta }$.  
According to LSDA+$U$ calculations \cite{junun}, $\underline{\Gamma}^{(i)}$ are given in the lab reference frame by
\begin{equation}
\underline{\Gamma}^{(1)} =
\left(
\begin{array}{ccc}
g & h & h \\
h& g & h \\
h& h &  g\\
\end{array} \right),
\end{equation}

\begin{equation}
\underline{\Gamma}^{(2)} =
\left(
\begin{array}{ccc}
0 & -j & j \\
j & 0 & -j \\
-j & j &  0\\
\end{array} \right),
\end{equation}
where $j=g-h$.
Transforming into the cycloidal reference frame,
\begin{equation}
\underline{U} \, \underline{\Gamma}^{(1)} \, \underline{U}^{-1} =
\left(
\begin{array}{ccc}
j & 0 & 0 \\
0 & j & 0 \\
0 & 0 &  g+2h \\
\end{array} \right),
\end{equation}

\begin{equation}
\underline{U} \, \underline{\Gamma}^{(2)} \, \underline{U}^{-1} =
\left(
\begin{array}{ccc}
0 & -\sqrt{3}j & 0 \\
\sqrt{3}j & 0 & 0 \\
0 & 0 &  0\\
\end{array} \right)
\end{equation}
for all three domains.  It follows that $C_{1\yq } = j$, $C_{1\zq } = g+2h$, and $C_2 = \sqrt{3}j$.
Therefore, $C_{1\yq }=C_2/\sqrt{3}$.

\section{ANI-induced polarizations} 

The perpendicular ANI-induced polarization $\vP^{{\rm ANI }}_{\perp }$ is 
associated with the dependence of the polarization direction $\vn $ on
an electric field.  The ANI energy is given by $V^{\rm ANI} =-K \sum_i (\vS_i \cdot \vn_i )^2$, where
\begin{equation}
\label{LSIA}
\vn_i = [\sin \theta_{iz} \cos \phi_{iz} , \sin \theta_{iz} \sin \phi_{iz} ,\cos \theta_{iz} ]
\end{equation} 
is the local single-ion ANI axis 
and $\theta_{iz} =\cos^{-1} (\vn_i \cdot \vz )$.  Consequently,
\begin{eqnarray}
\label{PAIz}
&&P^{{\rm ANI}}_z = -\frac{1}{N} \frac{dV^{{\rm ANI}}}{d E_z}
= -\frac{1}{N} \sum_i \Biggl\{
\frac{\partial V^{{\rm ANI}}}{\partial \theta_{iz}} \frac{\partial \theta_{iz} }{\partial E_z}\nonumber \\
&&+\frac{\partial V^{{\rm ANI}}}{\partial \phi_{iz}} \frac{\partial \phi_{iz} }{\partial E_z} \Biggr\}
-\frac{1}{N} \frac{\partial  V^{{\rm ANI}}}{\partial K} \frac{\partial K}{\partial E_z},
\end{eqnarray}
which must be evaluated in the limit $\vn_i \rightarrow \zp $.  Due to the rhombohedral crystal structure,
$\partial \theta_{ix}/\partial E_x =\partial \theta_{iy}/\partial E_y =\partial \theta_{iz}/\partial E_z$ where
$\theta_{ix} =\cos^{-1} (\vn_i \cdot \vx )$ and $\theta_{iy} =\cos^{-1} (\vn_i \cdot \vy )$.
Similar identities hold for $\phi_{i\alpha }$.  It follows that
\begin{eqnarray}
\label{PAIx}
&&P^{{\rm ANI}}_x = -\frac{1}{N} \frac{dV^{{\rm ANI}}}{d E_x}
= -\frac{1}{N} \sum_i \Biggl\{
\frac{\partial V^{{\rm ANI}}}{\partial \theta_{ix}} \frac{\partial \theta_{ix} }{\partial E_x}\nonumber \\
&&+\frac{\partial V^{{\rm ANI}}}{\partial \phi_{ix}} \frac{\partial \phi_{ix} }{\partial E_x} \Biggr\}
-\frac{1}{N} \frac{\partial  V^{{\rm ANI}}}{\partial K} \frac{\partial K}{\partial E_x},
\end{eqnarray}
\begin{eqnarray}
\label{PAIy}
&&P^{{\rm ANI}}_y = -\frac{1}{N} \frac{dV^{{\rm ANI}}}{d E_y}
= -\frac{1}{N} \sum_i \Biggl\{
\frac{\partial V^{{\rm ANI}}}{\partial \theta_{iy}} \frac{\partial \theta_{iy} }{\partial E_y}\nonumber \\
&&+\frac{\partial V^{{\rm ANI}}}{\partial \phi_{iy}} \frac{\partial \phi_{iy} }{\partial E_y} \Biggr\}
-\frac{1}{N} \frac{\partial  V^{{\rm ANI}}}{\partial K} \frac{\partial K}{\partial E_y}.
\end{eqnarray}
The first terms in Eqs.(\ref{PAIz}-\ref{PAIy}) then produce the $\xi_1= -4\sqrt{6}\, K  \, \partial \theta_{iz}/\partial E_z$ polarization perpendicular to $\zp $.  Because
$\partial \phi_{i\alpha } /\partial E_{\alpha }$ is modulated by $(-1)^{n_i}$, the second terms produce the 
$\xi_3= (2K/3) (-1)^{n_i}\, \partial \phi_{iz}/\partial E_z$ polarization perpendicular to $\zp$.  The final terms produce the $\xi_4 = 4\sqrt{3}\, \partial K/\partial E_z$ polarization 
$\vP^{{\rm ANI }}_{\parallel }$ along $\zp$.

In the lab reference frame, the perpendicular polarizations produced by the dependence of the polar and azimuthal angles $\theta_i $ and $\phi_i$ on the electric field $\vE $
are given respectively by 
\begin{equation}
\label{PANL1}
(\vP^{{\rm ANI (1)}}_{\perp })_ {\alpha }  = \frac{\xi_1 }{12\sqrt{3}N} \sum_{i ,\beta \gamma } \bigl( S_{i \alpha }
- S_{i\beta } \bigr) S_{i\gamma },
\end{equation}
\begin{equation}
\label{PANL2}
(\vP^{{\rm ANI (2)}}_{\perp })_{\alpha }  = \frac{\xi_3}{2N} \sum_{i, \beta \gamma \delta } (-1)^{n_i} \epsilon_{\alpha \beta \gamma } \bigl( S_{i\beta }-S_{i\gamma } \bigr)
S_{i\delta }.
\end{equation}
In the cycloidal reference frame, these polarizations are given by Eqs.(\ref{PAN1}) and (\ref{PAN2}).


\begin{thebibliography}{}
\bibitem{Hes} In previous work \cite{matsuda12}, the SW frequencies were scaled by $\sqrt{S(S+1)}$.  
This artificial introduction of quantum fluctuations made it awkward to obtain the correct SW frequencies of the 
canted AF phase, where the upper mode frequency is proportional to $2\mb H$ rather than $2\mb \sqrt{(S+1)/S}H$.
In the present work, the frequencies are scaled by $S$.  Consequently, the Heisenberg
interactions increase from $J_1=-4.50$ meV and $J_2 = -0.20$ meV to $J_1= -5.32$ meV and $J_2 = -0.24$ meV.
This rescaling was also applied to the quoted values of $D_1$ and $K$ in Ref.[\onlinecite{jeong14}].  Also keep in mind that
we have redefined $D_1$ to be $\sqrt{2}$ larger than in earlier work.
\bibitem{missing}  When $\vm =[0,0,1]$, a field-independent 
feature was reported in Ref.[\onlinecite{nagel13}] at 16.3 cm$^{-1}$.  Because it did not fit
into our mode scheme, this feature was difficult to understand.  A re-examination of the THz data reveals that 
the total area of the nearly-degenerate $\Psi_0/\Phi_1^{(1)}$ modes at zero field was erroneously reported too high 
($6.9 \pm 1.0$ cm$^{-2}$) by Ref.[\onlinecite{nagel13}].  The correct value is $3.5 \pm 0.3$ cm$^{-2}$ and there is no magnetic
field independent mode at 16.3 cm$^{-1}$.  Hence, it does not appear in Fig.4(a).
\bibitem{exc} It is questionable whether Ref.[\onlinecite{talbayev11}] actually observed the nearly-degenerate $\Psi_0/\Phi_1^{(1)}$ modes at zero field:  
the reported frequency of 18.4 cm$^{-1}$ is substantially higher than the 16.2 cm$^{-1}$ mode seen in Fig.4.  In Faraday 
measurements, Ref.[\onlinecite{nagel13}] also observed a feature in the absolute spectrum around 18 cm$^{-1}$ at zero field.
Since this feature does not appear in a magnetic field, it may be an interference fringe rather than an actual mode of the cycloid.
\bibitem{polar} A polar vector does not change sign under spatial inversion but an axial vector does.  Under spatial inversion,
\begin{eqnarray}
W_{1,x} &&= \frac{1}{N}\sum_{\vR_i ,\vR_j =\vR_i +a\vx }\vS_i \cdot \vS_j \nonumber \\
\rightarrow  && \frac{1}{N}\sum_{\vR_i ,\vR_j =\vR_i -a\vx }\vS_i \cdot \vS_j = W_{1,x} . \nonumber
\end{eqnarray}
Since similar relations hold for $W_{1,y}$ and $W_{1,z}$, $\vW_1$ is a polar vector.  On the other hand,
\begin{eqnarray}
W_{2,x} &&= \frac{1}{N}\sum_{\vR_i ,\vR_j =\vR_i +a\vx } (-1)^{n_i}\vS_i \cdot \vS_j \nonumber \\
\rightarrow && \frac{1}{N}\sum_{\vR_i ,\vR_j =\vR_i -a\vx } (-1)^{n_i}\vS_i \cdot \vS_j = - W_{2,x}  \nonumber
\end{eqnarray}
because $n_i = n_j +1$ for $\vR_i$ and $\vR_j = \vR_i - a\vx $.  Therefore, $\vW_2$ is an axial vector but $\zp \times \vW_2 $ is a polar vector.
\bibitem{mis} About 3\% of the light intensity has the wrong (perpendicular)
polarization.  This misalignment is more severe for the Voigt geometries 
($\vm = [1,-1,0]$ and [1,1,0]) than for the Faraday geometry ($\vm = [0,0,1]$).
\end{thebibliography}
\end{document}